\documentstyle[12pt,oldlfont]{article}

\expandafter\ifx\csname mathrm\endcsname\relax\def\mathrm#1{{\rm #1}}\fi

\makeatletter
\if@twoside \oddsidemargin -17pt \evensidemargin 00pt
\else \oddsidemargin 00pt \evensidemargin 00pt
\fi
\expandafter\ifx\csname mathrm\endcsname\relax\def\mathrm#1{{\rm #1}}\fi

\makeatother

\def\beq{\begin{equation}}
\def\eeq{\end{equation}}
\def\beqar{\begin{eqnarray}}
\def\eeqar{\end{eqnarray}}
\def\barr#1{\begin{array}{#1}}
\def\earr{\end{array}}
\def\bfi{\begin{figure}}
\def\efi{\end{figure}}
\def\btab{\begin{table}}
\def\etab{\end{table}}
\def\bce{\begin{center}}
\def\ece{\end{center}}
\def\nn{\nonumber}
\def\disp{\displaystyle}
\def\text{\textstyle}

\def\veps{\varepsilon}

\def\refse#1{\mbox{Sect.~\ref{#1}}}

\def\citere#1{\mbox{Ref.~\cite{#1}}}

\def\mathswitchr#1{\relax\ifmmode{\mathrm{#1}}\else$\mathrm{#1}$\fi}

\newcommand{\PW}{\mathswitchr W}
\newcommand{\PZ}{\mathswitchr Z}

\newcommand{\PH}{\mathswitchr H}

\newcommand{\Pt}{\mathswitchr t}

\newcommand{\PWO}{\mathswitchr {W^0}}

\def\mathswitch#1{\relax\ifmmode#1\else$#1$\fi}

\newcommand{\MW}{\mathswitch {M_\PW}}

\newcommand{\MZ}{\mathswitch {M_\PZ}}
\newcommand{\MH}{\mathswitch {M_\PH}}

\newcommand{\Mt}{\mathswitch {m_\Pt}}

\newcommand{\sw}{\mathswitch {s_\PW}}

\newcommand{\Ds}{D\hspace{-0.65em}/\hspace{0.25em}}

\newcommand{\alpz}{\alpha(\MZ^2)}
\newcommand{\bos}{{\mathrm{bos}}}
\newcommand{\fer}{{\mathrm{ferm}}}

\newcommand{\MVB} {{\mathrm{MVB}}}
\newcommand{\MVBB} {{\mathrm{MVB, bos}}}
\newcommand{\SM} {{\mathrm{SM}}}

\newcommand{\Dx}{\Delta x}
\newcommand{\Dy}{\Delta y}

\def\solid{\raise.9mm\hbox{\protect\rule{1.1cm}{.2mm}}}

\def\draftdate{\relax}
\def\mda{\relax}
\def\mua{\relax}
\def\mla{\relax}
\def\mpar#1{\relax}
\def\draft{
\def\draftdate{\today}
\def\mpar##1{\marginpar{\hbadness10000\sloppy\boldmath\bf##1}%
                      \typeout{marginpar: \noexpand##1}\ignorespaces}
\def\mda{\mpar{\hfil$\downarrow$\hfil}}
\def\mua{\mpar{\hfil$\uparrow$\hfil}}
\def\mla{\marginpar[\boldmath\hfil$\rightarrow$\hfil]%
                   {\boldmath\hfil$\leftarrow $\hfil}%
                    \typeout{marginpar: $\leftrightarrow$}\ignorespaces}
}

\makeatletter

\def\eqnarray{\stepcounter{equation}\let\@currentlabel=\theequation
\global\@eqnswtrue
\global\@eqcnt\z@\tabskip\@centering\let\\=\@eqncr
$$\halign to \displaywidth\bgroup\hskip\@centering
  $\displaystyle\tabskip\z@{##}$\@eqnsel&\global\@eqcnt\@ne
  \hskip 2\arraycolsep \hfil${##}$\hfil
  &\global\@eqcnt\tw@ \hskip 2\arraycolsep $\displaystyle\tabskip\z@{##}$\hfil
   \tabskip\@centering&\llap{##}\tabskip\z@\cr}
\def\appendix{\par
 \setcounter{section}{0} \setcounter{subsection}{0}
 \def\thesection{\Alph{section}}}

\makeatother

\begin{document}
\thispagestyle{empty}
\def\thefootnote{\fnsymbol{footnote}}
\setcounter{footnote}{1}
\hfill \begin{minipage}[t]{4cm}BI-TP 94/31\\
hep-ph/9406378\end{minipage}
\vspace*{2cm}
\begin{center}
{\Large\bf On the Role of the Higgs Mechanism}   \\
{\Large\bf in Present Electroweak Precision Tests%
\footnote{Supported by the Bundesministerium f\"ur Forschung
und Technologie, Bonn, Germany.}}%
\end{center}
\vspace*{0.3cm}
\begin{center}
S.\ Dittmaier, C.\ Grosse-Knetter and D.\ Schildknecht
\\[.3cm]
University of Bielefeld, Department of Theoretical Physics, \\[.3cm]
33501 Bielefeld, Germany
\end{center}
\vspace*{3cm}
\section*{Abstract}
Based on the observables $\MW$, $\Gamma_l$, $\bar\sw^2(\MZ^2)$,
we evaluate the parameters $\Delta x, \Delta y$ and $\varepsilon$
at one-loop level within an electroweak massive vector-boson
theory, which does not employ the Higgs mechanism.
The theoretical results are consistent with the experimental ones on
$\Delta x$, $\Delta y$, $\varepsilon$. The theoretical prediction for
$\Delta y$ coincides with the standard-model one (apart from numerically
irrelevant terms which vanish for $\MH\to\infty$). Non-renormalizability
only affects $\Delta x$ and $\varepsilon$, which differ from the
standard-model results by the replacement $\log\MH\to\log\Lambda$ for a
heavy Higgs mass, $\MH$ (where $\Lambda$ denotes an effective UV
cut-off).
\vfill
June 1994 \hfill

\def\thefootnote{\arabic{footnote}}
\setcounter{footnote}{0}
\clearpage

\section{Introduction}

Our recent analysis \cite{DITT} of the electroweak precision data
\cite{LEP} was based on an effective Lagrangian \cite{BIL}
which allows for
$SU(2)$ breaking via non-vanishing values of the
parameters $\Delta x , \Delta y$ and $\varepsilon$\footnote{
The parameter
$\Delta x$ quantifies global $SU(2)$ violation via
$M^2_{W^\pm} \equiv (1 + \Delta x) M^2_\PWO$,
while $\Delta y$ and $\varepsilon$ quantify $SU(2)_L$ violation in
vector-boson couplings to leptons
$4\sqrt 2 G_\mu M^2_{W^\pm}\equiv
g^2_{W^\pm} (0) \equiv (1 + \Delta y) g^2_\PWO (\MZ^2)$,
and via mixing
${\cal L}_{{\rm mix}} \equiv (e(\MZ^2)/ g_\PWO(\MZ^2))  (1 - \varepsilon )
A_{\mu\nu} W^{\mu\nu}_3$.
}.
The
interpretation of the results of this analysis is twofold:
\renewcommand{\labelenumi}{(\roman{enumi})}
\begin{enumerate}
\item
Independently of any model for the parameters
$\Delta x, \Delta y, \varepsilon$ the
very small empirical values of these parameters
{\it strongly restrict violations of the
$SU(2)_L \times U(1)_Y$ symmetry} which
is at the root \cite{GL} of present-day (standard)
electroweak theory \cite{WS}.
\item
Comparison with the one-loop predictions of the standard
$SU(2)_L \times U(1)_Y$ electroweak theory for
$\Delta x, \Delta y, \varepsilon$ provides a test of this
theory at one-loop level. Our analysis \cite{DITT} has revealed that
the most recent experimental results have
reached a {\it precision} which for the first time allows to
{\it test the
theory beyond pure fermion-loop effects}.
\end{enumerate}

In the present note, we examine the implications of the results of
the electroweak
precision tests with respect to the underlying
notions of spontaneous symmetry
breaking and the
Higgs mechanism.

Based on the observables $\MW$, $\Gamma_l$, $\bar\sw^2(\MZ^2)$,
in \citere{DITT}, the one-loop results for $\Delta x,
\Delta y, \varepsilon$ were systematically and explicitly
represented as a sum of
pure fermion-loop (vacuum-po\-la\-ri\-za\-tion) corrections
and the remaining bosonic ones,
\beq
a = a_\fer + a_\bos,\qquad{\rm with} \quad
a = \Delta x, \Delta y, \varepsilon .
\label{a}
\eeq
The fermion part in (\ref{a}) depends on $\alpha(\MZ^2)$, $s^2_0$
and the mass of the top quark, $\Mt$,
\beq
a_\fer = a_\fer(\alpz,s^2_0,\Mt^2/\MZ^2),
\label{b}
\eeq
where $\alpha (\MZ^2) \cong 1/129$ denotes the electromagnetic
coupling at the scale $\MZ^2$ \cite{JEG}, and
\beq
s^2_0 (1 - s^2_0 ) \equiv s^2_0 c^2_0 \equiv
\frac{\pi \alpha (\MZ^2)}{\sqrt 2 G_\mu \MZ^2}.
\label{c}
\eeq
The fermion-loop contributions in (\ref{a})
will not be our primary concern in the present investigation.
We concentrate on the bosonic contribution in (\ref{a}), which
in general depends on the trilinear and
quadrilinear interactions of the vector bosons among themselves
and the mass and couplings of the Higgs scalar.
More specifically, one has to discriminate between
$\Delta x_\bos$ and $\varepsilon_\bos$ on the one hand
and $\Delta y_\bos$ on the other hand.
Whereas $\Delta x_\bos$ and $\varepsilon_\bos$
depend on $\log\MH$ for a heavy Higgs,
no such dependence is present in $\Delta y_{\rm bos}$.
Moreover, relative to present (and future) experimental uncertainty,
the bosonic contributions in the full expressions (\ref{a})
for $\Delta x$ and $\varepsilon$ are small, i.e.\ in
reasonable approximation the bosonic contributions
to $\Delta x$ and $\varepsilon$ in (\ref{a}) can be neglected
in the standard electroweak theory,
\beq
\Delta x \cong \Delta x_\fer, \qquad
\varepsilon \cong \varepsilon_\fer.
\label{f}
\eeq
In contrast, $\Delta y$ in (\ref{a})
receives a substantial bosonic contribution,
\beq
\Delta y_\bos \cong 14 \cdot 10^{-3},
\label{h}
\eeq
whose magnitude is by no means negligible. Indeed, the
data on $\Delta y$
require a contribution beyond fermion loops,
which is consistent with the standard prediction for
$\Delta y_\bos$ in (\ref{h}).

The mentioned properties of $\Delta x, \Delta y, \varepsilon$ in the
standard electroweak theory can be clearly identified in Figs.~4-6
of \citere{DITT}, which also contain the comparison with the
experimental results \cite{LEP}.

We note that $\Delta y_\bos$ is the only one of the
three parameters, $\Delta x, \Delta y, \varepsilon$, which
involves the radiative corrections to $\mu^\pm$ decay (the
$W^\pm$-fermion-vertex corrections, in particular)
at one-loop level. The fact that $\Delta y_\bos$ is
practically independent of $\MH$ indicates that for
the finiteness of the (one-loop) radiative corrections
to $\mu^\pm$-decay the
existence of the Higgs particle is irrelevant.

More generally, it is a relevant and importent question to be asked,
{\it to what extent the
precision-test parameters
$\Delta x, \Delta y, \varepsilon$ can
be predicted at one-loop level without employing the Higgs mechanism}. It is
indeed well-known that the standard-model interactions between
fermions and
vector bosons and the interactions among the vector bosons themselves
can credibly be derived \cite{HS,SCH} in (non-renormalizable)
massive vector-boson theories with mixing
in the neutral-boson sector. In other words, the $\gamma
W^3$-current-mixing or the $BW^3$-mass-mixing ansatz, not being
based on the notion of spontaneous symmetry breaking, lacks the
Higgs scalar and, consequently, renormalizability, but otherwise the
Lagrangian coincides with the standard electroweak theory \cite{WS}
in its unitary gauge with the Higgs scalar field omitted.
As a consequence, the tree-level predictions for fermion-fermion interactions
in the massive vector-boson theory agree with the ones from the standard
electroweak theory.

In the present note, we analyze $\Delta x, \Delta y, \varepsilon$
within the {\em massive vector-boson theory at
one-loop level.\/}
Not entirely unexpected, we will find that the one-loop prediction
for $\Dy_{\rm bos}$ in the massive vector-boson theory
also yields (\ref{h}). As a consequence, the parameter $\Dy$, the
bosonic contribution to which can be experimentally determined, does not
allow to discriminate between the standard theory based on the
Higgs mechanism and the Higgs-less massive-vector-boson theory. The
parameters $\Dx_{\rm bos}$ and $\veps_{\rm bos}$ in the
massive-vector-boson theory contain a logarithmic divergence quantified
by a $\log \Lambda$ cut-off dependence, obviously related to the
fact that the massive electroweak vector-boson theory is incomplete.
For a huge range in $\Lambda$ the bosonic contributions to
$\Dx$ and $\veps$ are negligible, the approximation (\ref{f}) in this
sense thus remaining valid in the massive vector boson theory. This
is related to the correspondence $\log \Lambda \leftrightarrow
\log \MH$ between the cut-off dependence of the Higgs-less model and the
dependence on the (heavy) Higgs mass in the electroweak standard model,
frequently elaborated upon \cite{APBE,HEMO} from a
somewhat different point of view.

In \refse{MVB}, we briefly remind ourselves
of how standard electroweak interactions
of massive vector bosons, photons
and fermions are reconstructed without employing the
Higgs mechanism. In a second step, the Stueckelberg transformation
\cite{STU} generalized to massive Yang-Mills fields
is used \cite{KUN} to
restore
$SU(2)_L \times U(1)_Y$ invariance of the basic Lagrangian.
Accordingly, we are able to introduce gauge-fixing terms as usual.
After this preparation it amounts to a simple step to
find the explicit and complete expressions for the three
parameters $\Delta x$, $\Delta y$, $\varepsilon$
in \refse{onel} in the Higgs-less theory. The corresponding
comparison with the experimental data is seen to be
implicitly contained in Figs.~4-6 of Ref.~\cite{DITT}.
Conclusions will be
drawn in \refse{concl}.

\section{Gauge-invariant Electroweak Massive Vector-Boson Theory}
\label{MVB}

We start with the group $SU(2)_L \times U(1)_Y$ \cite{GL}.
Introducing the weak-isospin $W^\pm, \PWO$ triplet $W^i_\mu (x)$
$(i = 1,2,3)$
and a hypercharge singlet, $B_\mu (x)$, coupled to a
left-handed doublet and a right-handed singlet, we have\footnote{
Without loss of generality in this context, we restrict
ourselves to a single fermion doublet
.}
\beq
{\cal L} = - \frac{1}{2} {\rm tr} (W^{\mu\nu} W_{\mu\nu} ) -
\frac{1}{4} B^{\mu\nu} B_{\mu\nu}
+ i (\overline{\Psi}_L \Ds \Psi_L + \overline{\Psi}_R \Ds \Psi_R ).
\label{j}
\eeq
with
\beq
W_\mu = W^i_\mu \frac{\tau_i}{2}, \qquad
W_{\mu\nu} = \partial_\mu W_\nu - \partial_\nu W_\mu + i g
[W_\mu , W_\nu ]
\label{k}
\eeq
and
\beqar
\Psi_{R} = \frac{1}{2} ( 1+\gamma^5)\Psi , &\quad&
D_\mu \Psi_R = (\partial_\mu + \frac{i}{2} g^\prime Y_R B_\mu)
\Psi_R ,
\nonumber \\
\Psi_{L} = \frac{1}{2} ( 1-\gamma^5)\Psi , &\quad&
D_\mu \Psi_L = (\partial_\mu - i g W_\mu + \frac{i}{2}
g^\prime Y_L B_\mu ) \Psi_L .
\label{l}
\eeqar
The weak hypercharges, $Y_{L/R}$, and the baryon number, $B$,
and lepton number, $L$, are related as follows
\beq
Y_L = B-L, \qquad Y_R = \tau_3+B-L.
\label{m}
\eeq
The Lagrangian is invariant under {\em local\/}
$SU(2)_L \times U(1)_Y$
transformations,
\beqar
W_\mu &\rightarrow &
\rlap{$
SW_\mu S^\dagger - \frac{i}{g} S \partial_\mu
S^\dagger, \qquad
S = \exp \left( \frac{i}{2} g \alpha_i \tau_i \right),
$} \hspace{20em}
\alpha_i= \alpha_i(x),
\nonumber \\
B_\mu &\rightarrow &
\rlap{$
B_\mu - \partial_\mu\beta,
$} \hspace{20em}
\beta=\beta(x),
\label{z}
\eeqar
and
\beq
\Psi_L \rightarrow S \exp\left( \frac{i}{2} g^\prime Y_L
\beta \right) \Psi_L , \qquad
\Psi_R \rightarrow \exp\left( \frac{i}{2} g^\prime Y_R
\beta \right) \Psi_R .
\label{o}
\eeq
Boson masses
are introduced in (\ref{j}) in such a manner that local
{\it electromagnetic} gauge invariance, $U(1)_{em}$,
\beq
B_\mu \rightarrow B_\mu + \frac{1}{g^\prime} \partial_\mu \chi,
\qquad
W_{3\mu} \rightarrow W_{3\mu} + \frac{1}{g} \partial_\mu \chi,
\qquad
W_\mu^\pm \rightarrow \exp\left(\mp ie\chi\right)W_\mu^\pm
\label{am}
\eeq
is preserved. The most general mass-mixing term is then given by
\beq
{\cal L}_{{\rm bos, mass}} = M^2_W \left[ W_\mu - \frac{1}{2}
\frac{g^\prime}{g} B_\mu \tau_3 \right]^2 .
\label{n}
\eeq
Diagonalization of the mass term in the Lagrangian
implies Weinberg's mass relations \cite{WS} and couplings of the vector
bosons, i.e.\ the standard model without Higgs boson.
In addition, one has to introduce the fermion mass term
\beq
{\cal L}_{\rm ferm, mass}=-\bar{\Psi}_L M_{\rm F} \Psi_R
-\bar{\Psi}_R M_{\rm F} \Psi_L\qquad \mbox{with}\qquad M_{\rm F}
=\left(\begin{array}{cc}m_1&0\\0&m_2\end{array}\right).
\label{fermmass}
\eeq

These mass terms
as they stand, seem to break local $SU(2)_L \times
U(1)_Y$ symmetry. Local $SU(2)_L\times U(1)_Y$ symmetry is
immediately restored, however, by introducing \cite{KUN}
auxiliary fields
$\varphi_i(x)$ \'a la Stueckelberg \cite{STU} via
the substitutions
\beq
W_\mu \rightarrow U^\dagger W_\mu U - \frac{i}{g}
U^\dagger \partial_\mu U, \qquad
B_\mu \rightarrow B_\mu , \qquad
\Psi_R \rightarrow \Psi_R, \qquad
\Psi_L \rightarrow U^\dagger\Psi_L,
\label{p}
\eeq
where
\beq
U \equiv \exp \left( \frac{i}{2} \frac{g}{\MW} \varphi_i \tau_i
\right).
\label{ap}
\eeq
The substitutions leave the kinetic and interaction
terms given in (\ref{j}) invariant, while the mass
terms (\ref{n}), (\ref{fermmass}) take the form
\beq
{\cal L}_{\rm mass}=
\frac{M^2_W}{g^2} {\rm tr}\left[ (D_\mu U)^\dagger (D^\mu U) \right]
-\bar{\Psi}_L U M_{\rm F} \Psi_R-\bar{\Psi}_R M_{\rm F} U^\dagger \Psi_L,
\label{q}
\eeq
where the covariant derivative is defined by
\beq
D_\mu U \equiv \partial_\mu U + ig W_\mu U - \frac{i}{2} g^\prime
U \tau_3 B_\mu .
\label{r}
\eeq
The $SU(2)_L \times U(1)_Y$ gauge transformation (\ref{z}), (\ref{o}) and
\beq
U \rightarrow S U \exp \left( \frac{-i}{2}
g^\prime \beta \tau_3 \right)
\label{s}
\eeq
assure invariance of the full electroweak Lagrangian including
the vector-boson mass terms%
\footnote{It is interesting to note that
an arbitrary $SU(2)_L\times U(1)_Y$ gauge transformation (\ref{z}),
(\ref{o}), (\ref{s}) applied to the fields in the
Lagrangian (\ref{j}), (\ref{q}) formulated \'a la Stueckelberg,
acts on the physical fields in
the original electroweak massive-vector-boson theory
(\ref{j}), (\ref{n}), (\ref{fermmass}) simply
as an electromagnetic $U(1)_{em}$ gauge transformation
given by (\ref{am}).}.
A suitable gauge transformation yields $U = 1$ and takes us back to
the original Lagrangian which is thus identified as a $SU(2)_L\times U(1)_Y$
gauge-invariant massive vector-boson theory with mixing in the neutral sector,
represented in a specific gauge.

For comparison with the standard electroweak theory \cite{WS} based
on the Higgs-Kibble mechanism \cite{HK}, we expand the exponential in
(\ref{ap}). Dropping ${\cal O} (\varphi^2_i)$ terms, upon
substitution in (\ref{q}), one finds the kinetic term of the
Higgs-field part of the standard theory under the restriction of
$H(x) = 0$, where $H(x)$ denotes the additional physical
field appearing in
the complex Higgs doublet
\beq
\Phi=\frac{1}{\sqrt{2}}((v+H){\bf 1}+i\tau_i\varphi_i)
\label{higgs}
\eeq
of the standard theory, i.e.
\beq
\frac{M^2_W}{g^2} {\rm tr} \left[ (D_\mu U )^\dagger (D^\mu U)
\right] = \frac{1}{2} {\rm tr} \left.\left[ (D_\mu \Phi )^\dagger
(D^\mu \Phi )\right]\right\vert_{H=0} + {\cal O} (\varphi^2_i).
\label{v}
\eeq
The higher powers of $\varphi_i$ yield interactions of
arbitrarily many scalar fields, not present in the standard theory.
For the specific class of four (light) fermion processes at one-loop
level, the Feynman diagrams and the corresponding amplitudes are the
same as in the standard model
except that the graphs containing an internal Higgs field are
obviously absent and some of the tadpole-like graphs with
vector-vector-scalar-scalar couplings slightly change\footnote{In
particular, the modifications of the tadpole-like graphs cancel
in the determination of $\Dx$, $\Dy$, $\veps$.}.

The $SU(2)_L \times
U(1)_Y$ gauge-invariant Lagrangian
(\ref{j}), (\ref{q})
arrived at by applying the generalized
Stueckelberg formalism coincides with the Lagrangian of the so-called
$SU(2)_L \times U(1)_Y$ gauged non-linear $\sigma$-model\footnote{
Even though this terminology correctly identifies the
underlying mathematics, it is somewhat unfortunate as it
associates the name of a {\it physical} theory exactly with
{\it unphysical} degrees of freedom appearing therein.}
which has been frequently
investigated with respect to its leading (logarithmic) one-loop divergences
\cite{APBE}, one-loop divergences plus finite terms \cite{HEMO}
and two-loop divergences \cite{BIVE}. In those
investigations, the starting point is the search for the
leading dependence on the Higgs mass, $\MH$, at one-loop level
in the Weinberg Salam model \cite{WS}. The necessary calculations were
simplified by treating the non-linear $\sigma$-model at one-loop
level and by subsequently using the (conjectured) identity of the
resulting $\log \Lambda$ divergencies
with the $\log \MH$ terms in the Weinberg Salam model.

In contrast, in the present work,
the auxiliary Stueckelberg fields have been introduced
as a technical device without reference to the Higgs mechanism
in order to perform
calculations in the massive vector boson theory
(\ref{j}), (\ref{n}), (\ref{fermmass})
in an arbitrary $R_\xi$ gauge. Accordingly,
the one-loop expression
for the parameters $\Delta x, \Delta y, \varepsilon$,
(with respect to divergent as well as finite terms) of Sect.~3 are obtained
as consequences of the massive vector-boson theory with mixing specified
by the Lagrangian (\ref{j}), (\ref{n}), (\ref{fermmass}).

\section{One-loop results for $\Delta x, \Delta y, \varepsilon$}
\label{onel}

Since the massive vector-boson theory described above is
non-renormalizable, the one-loop corrections to the observables $\MW,
\Gamma_l , \bar s_W^2 (\MZ^2)$ remain UV divergent even after
the on-shell renormalization of the particle masses and the
identification of $e(0)$ in the Thomson limit\footnote{
The Weinberg angle $\theta_{\rm W}$ is fixed by $\cos\theta_{\rm W}\equiv
\frac{\MW}{\MZ}$ in the complete on-shell renormalization scheme.}.
Instead of introducing a further renormalization constant (in order
to render the vector-boson--fermion coupling finite), we rewrite
the divergencies via
\beq
\frac{2}{4-d} - \gamma_E + \log 4\pi + \frac{5}{6} + \log \mu^2
\;\longrightarrow\; \log \Lambda^2,
\label{w}
\eeq
where $d$ and $\mu$ denote the space-time dimension and the
reference mass of dimensional regularization, respectively
($\gamma_E =$ Euler's constant). Consequently, the mass scale
$\Lambda$ plays the role of an effective `cut-off parameter' or a
threshold for new physics \cite{VELT}.

In this way, we are able to
calculate the bosonic one-loop contributions to the
parameters $\Delta x, \Delta y, \varepsilon$ within the massive
vector-boson (\MVB) theory in complete analogy to the standard-model
(SM) evaluation reported in \citere{DITT}.
Of course, the pure fermionic corrections to $\Delta x$, $\Delta y$,
$\varepsilon$ coincide in both models at one-loop.
The resulting values for the bosonic contributions to the
parameters $a_{{\rm MVB}}$ $(a =
\Delta x, \Delta y, \varepsilon)$ depend on the following variables
\beqar
\Delta x_\MVBB &=& \Delta x_\MVBB \left(\alpz,
s^2_0 , \log (\Lambda^2 / \MZ^2)\right), \nn\\
\Delta y_\MVBB &=& \Delta y_\MVBB\left(\alpz,s^2_0\right),\nn\\
\varepsilon_\MVBB &=& \varepsilon_\MVBB \left(\alpz,
s^2_0 , \log (\Lambda^2 / \MZ^2)\right).
\label{dxyemvb}
\eeqar
Rather than explicitly giving the full expressions for the
quantities (\ref{dxyemvb}), it is convenient and sufficient to
present them implicitly by relating them to the results of the
standard electroweak theory given in \citere{DITT}. Within the
present context these results can be reconstructed by adding the
contributions of the diagrams with internal Higgs lines to the
results of the massive vector-boson theory. One then obtains
\beqar
\Delta x_{{\rm SM,bos}} &=& \Delta x_\MVBB
+\frac{3\alpz}{16\pi c^2_0}\log\left(\frac{\Lambda^2}{\MH^2}\right)
+ \Delta x_{{\rm SM ,bos}}
(rem), \nonumber \\
\Delta y_{{\rm SM,bos}} &=&
\rlap{$\disp \Delta y_\MVBB$}
\phantom{\Delta x_\MVBB
+\frac{3\alpz}{16\pi c^2_0}\log\left(\frac{\Lambda^2}{\MH^2}\right)}
+ \Delta y_{{\rm SM , bos}}
(rem) ,  \nonumber \\
\varepsilon_{{\rm SM, bos}} &=&
\rlap{$\disp
\varepsilon_\MVBB
+\frac{\alpz}{48\pi s^2_0}\log\left(\frac{\Lambda^2}{\MH^2}\right)$}
\phantom{\Delta x_\MVBB
+\frac{3\alpz}{16\pi c^2_0}\log\left(\frac{\Lambda^2}{\MH^2}\right)}
+\varepsilon_{{\rm SM , bos}} (rem).
\label{x}
\eeqar
{}From \citere{DITT}, we recall that the remainder terms $a_{\SM,\bos} (rem)
(a = \Dx, \Dy, \veps)$ in (\ref{x}) are defined as the difference
between the asymptotic $(\MH \rightarrow \infty)$ and the
exact $\MH$ dependence, i.\ e.\
\beq
a_{\SM,\bos} (rem) = {\cal O} \left( \frac{\MZ^2}
{\MH^2} \log \left(\MH^2/\MZ^2\right) \right)
\mathop{\longrightarrow}\limits_{\MH \rightarrow\infty} 0.
\label{y}
\eeq
These contributions in (\ref{x}), already for values of $\MH$ not very
far above $\MZ$, are irrelevant in comparison with (present and
future) experimental accuracy (see e.\ g.\ Tab. 1 in
\citere{DITT}) and can be disregarded in the ensuing discussions of
the result (\ref{x}).

In \citere{DITT} we noted that a non-vanishing bosonic contribution,
$\Dy_{\rm bos} \not= 0$, is required by the most recent data.
The magnitude of $\Dy_\bos$ required by the data was found to be
consistent with the standard prediction. According to (\ref{x}),
upon disregarding the remainder term, $\Dy_{SM, bos} ({\rm rem})$,
we find that $\Dy_{\rm MVB, bos}$ is equal to $\Dy_{\rm SM, bos}$.
Consequently, even utmost precision on $\Dy_{\rm bos}$ does not yield
a direct test of the Higgs mechanism. Theoretically, the identity of
$\Dy_{\rm MVB, bos}$ and $\Dy_{\rm SM, bos}$ may be expected from
the fact that $\Dy_{\rm SM, bos}$does not contain a $\log \MH$ term. It
is nevertheless gratifying to see the result emerging in the electroweak
massive vector-boson theory {\it without reference to the Higgs
mechanism}.

With respect to $\Dx$ and $\veps$, in \citere{DITT}, we found that
the precision data do {\it not} require bosonic contributions to
$\Dx$ and $\veps$. The data are consistent with relation (\ref{f}).
Accordingly, and
in contrast to the case of $\Dy$, for $\Dx$ and $\veps$, {\it any
theoretical ansatz with sufficiently suppressed bosonic contributions}
will be consistent with the precision data. Returning to (\ref{x}), upon
disregarding the remainders, $\Dx_{\SM,\bos}(rem)$ and
$\veps_{\SM,\bos}(rem)$,
we conclude that standard electroweak theory and massive vector boson
theory coincide for $\Lambda \equiv \MH$,
\beqar
\Dx_{\rm SM, bos} &\cong& \Dx_{\rm MVB}
\left( \alpz, s_0^2, \log (\Lambda^2/\MZ^2) \right)
\bigg|_{\Lambda = \MH},  \nonumber \\
\veps_{\rm SM, bos} &\cong& \veps_{\rm MVB}
\left( \alpz, s_0^2, \log (\Lambda^2/\MZ^2) \right)
\bigg|_{\Lambda = \MH},
\label{ax}
\eeqar
and both are equally suppressed provided $\Lambda$ in the massive
vector-boson theory is chosen in the
interval usually accepted for the Higgs-mass, $\MH$. While both, the
standard electroweak theory and the electroweak
massive-vector-boson ansatz thus appear to be consistent with the data
on $\Dx$ and $\veps$, it is to be kept in mind that the
identification%
\footnote{It should be noted that a constant term $\frac{5}{6}$
has been absorbed into $\log\Lambda^2$ in (\ref{w}) which thus does not
simply correspond to $2/(4-d)+\gamma_E+\log4\pi+\log\mu^2$ as conjectured
in \citere{APBE}. There are additional finite contributions
\cite{HEMO} which can in general not be completely absorbed by
a single universal cut-off parameter.}
\beq
\log \Lambda \;\leftrightarrow\; \log \MH
\label{ay}
\eeq
provides the only explicitly known genuine physical realization of the
cut-off $\Lambda$.
Introducing the Higgs particle guarantees \cite{THOOFT} perturbative
renormalizability (provides $\MH \leq 0.8$ TeV). On the
other hand, it cannot be excluded by present experimental
information that nature choses in fact an alternative
mechanism in order to remove the logarithmic divergences
appearing in the massive-vector-boson theory at one-loop order.
Detailed empirical information on the existence of the Higgs particle
and, more generally, on vector-boson scattering will be indispensable.

The relation between the logarithmic Higgs-mass dependence (including
the next-to-leading constant contribution) in the standard model with a
heavy Higgs boson and the corresponding UV divergencies appearing in the
non-linear $\sigma$-model was recently investigated in \citere{HEMO}.
Using
\beq
\Dx=\veps_{N1}-\veps_{N2},\qquad\Dy=-\veps_{N2},\qquad \veps=-\veps_{N3},
\label{alt}
\eeq
between our parameters and the ones of \citere{ALT}, upon taking into
account (\ref{w}), we find complete
agreement with respect to the cut-off terms in (\ref{x}).

\section{Conclusion}
\label{concl}

In summary, the parameter $\Delta y$, the only one to which
contributions beyond fermion loops are required by the
experimental data, assumes identical one-loop values in the
massive vector-boson theory and the standard model with a heavy Higgs
boson based on the
Higgs mechanism. Although utmost experimental precision on this
parameter does not teach us anything on the Higgs mechanism,
the empirical information is extremely important with respect
to testing other ingredients of the theory, such as the interactions
of the vector bosons among each other. In $\Delta x$ and
$\varepsilon$, massive vector-boson theory and standard results at
one-loop level are related to each other via the replacement
$\log \Lambda \leftrightarrow \log \MH$
(apart from numerically irrelevant terms which
vanish for $\MH\to \infty$).
Both
theories yield
exceedingly small bosonic effects in $\Delta x$ and $\varepsilon$.

The overall-conclusion comes without surprise. The
high-precision data on $\Gamma_l$, $\bar\sw^2(\MZ^2)$, $\MW$ are
consistent with the standard theory based on breaking an
underlying $SU(2)_L \times U(1)_Y$ local symmetry via the
Higgs mechanism. The data can also be reproduced,
however, by evaluating the massive vector-boson theory at
one-loop level. The issue of mass generation via the Higgs
mechanism will remain an open one as long as more direct empirical
evidence for the Higgs scalar particle is lacking.

\end{document}